\documentclass[aps,prl,superscriptaddress,twocolumn,showpacs]{revtex4}
\usepackage{amsmath}
\usepackage{graphicx,epsfig}

\renewcommand{\v}[1]{{\bf #1}}
\newcommand{\be}{\begin{equation}}
\newcommand{\ee}{\end{equation}}
\newcommand{\nn}{\nonumber\\}

\newcommand{\ba}{\begin{eqnarray}}
\newcommand{\ea}{\end{eqnarray}}

\newcommand{\cbar}{c^+}

\begin{document}

\title{Skyrmions and Anomalous Hall Effect in a Dzyloshinskii-Moriya
Spiral Magnet}
\author{Su Do Yi}
\affiliation{Department of Physics, BK21 Physics Research Division,
Sungkyunkwan University, Suwon 440-746, Korea}
\author{Shigeki Onoda}
\affiliation{Condensed Matter Theory Laboratory, RIKEN, 2-1,
Hirosawa, Wako 351-0198, Japan}
\author{Naoto Nagaosa}
\affiliation{Department of Applied Physics, The University of Tokyo,
7-3-1, Hongo, Bunkyo-ku, Tokyo 113-8656, Japan} \affiliation{Cross
Correlated Materials Research Group, Frontier Research System, Riken,
2-1 Hirosawa, Wako, Saitama 351-0198, Japan}
\author{Jung Hoon Han}
\email[Electronic address:$~~$]{hanjh@skku.edu}
\affiliation{Department of Physics, BK21 Physics Research Division,
Sungkyunkwan University, Suwon 440-746, Korea}
\date{\today}

\begin{abstract}
Monte Carlo simulation study of a classical spin model with
Dzylosinskii-Moriya interaction and the spin anisotropy under the
magnetic field is presented. We found a rich phase diagram
containing the multiple spin spiral (or skyrme crystal) phases of
square, rectangular, and hexagonal symmetries in addition to the
spiral spin state. The Hall conductivity $\sigma_{xy}$ is calculated
within the $sd$ model for each of the phases. While $\sigma_{xy}$ is
zero in the absence of external magnetic field, applying a field
strength $H$ larger than a threshold value $H_c$ leads to the
simultaneous onset of nonzero chirality and Hall conductivity. We
find $H_c = 0$ for the multiple spin spiral states, but $H_c > 0$
for a single spin spiral state regardless of the field orientation.
Relevance of the present results to MnSi is discussed.
\end{abstract}


\maketitle

Spiral magnets due to the Dzyloshinskii-Moriya (DM) interaction are
attracting recent intense attention. MnSi is one such prominent
example of a metallic chiral ferromagnet whose
magnetic\cite{ishikawa,neutron-on-MnSi,partial-order} and
transport\cite{transport,ong1,ong2} properties have been thoroughly
studied over several decades. 
Also the spiral magnet is a promising laboratory to study the
important issue of the relationship between the nontrivial spin
textures and the quantum electronic transport, since the
non-collinear spin configurations can be easily realized in such
magnets. Indeed, Lee \textit{et al}. found the novel anomalous Hall
effect (AHE) in a certain region of the pressure-temperature phase
diagram of MnSi recently\cite{ong1,ong2}. Several calculations have
pointed out a close connection between the spin chirality derived
from the non-collinear spin ordering, the Berry phase associated with
it, and the anomalous Hall effect which arises as a
consequence\cite{ye,geller,bruno,review}.

The purpose of this paper is twofold. First, we address the possible
phase diagram of a spiral magnet from a microscopic model that
includes the DM interaction and spin anisotropy. The phase diagram in
the plane of spin anisotropy-magnetic field including the multiple
spin spiral or Skyrmion crystal phase of square, rectangular, and
hexagonal symmetries is revealed. The anomalous Hall conductivity is
calculated for a model of conduction electrons coupled to the spins
via the double exchange interaction, and the relevance of Skyrmion
spin texture on the Hall transport is studied. In this way, a clear
and consistent connection is drawn between the underlying topological
spin structure, and its manifestation in an anomalous Hall transport.

%

A continuum Hamiltonian written by Bak and Jensen for a prototypical
chiral magnet MnSi some years ago\cite{bak} is adapted to a lattice
spin model consisting of the ferromagnetic exchange ($J$), DM
interaction ($K$), anisotropy ($A_1$ and $A_2$), and the Zeeman
energy ($H$):

\ba && H_S = -J \sum_{\v r} \v S_{\v r} \cdot (\v S_{\v
r\!+\!\hat{x}} \!+\!\v S_{\v r\!+\!\hat{y}}\!+\!\v S_{\v
r\!+\!\hat{z}}) \nn
&& - K \sum_{\v r}( \v S_\v r \!\times\! \v S_{\v r\!+\!\hat{x}}
\!\cdot \!\hat{x} \!+\! \v S_{\v r} \!\times\! \v S_{\v
r\!+\!\hat{y}} \!\cdot \! \hat{y} \!+\!\v S_{\v r} \!\times\! \v
S_{\v r\!+\!\hat{z}} \!\cdot \! \hat{z}) \nn
&& ~~~~ +A_1 \sum_{\v r} \Bigl( (S^x_{\v r} )^4 \!+\! (S^y_{\v r} )^4
\!+\! (S^z_{\v r} )^4 \Bigr) \nn
&& - A_2 \sum_{\v r} \Bigl(S^x_{\v r} S^x_{\v r\!+\!\hat{x}} \!+\!
S^y_{\v r} S^y_{\v r\!+\!\hat{y}} \!+\! S^z_{\v r} S^z_{\v
r\!+\!\hat{z}} \Bigr) - \v H \cdot \sum_{\v r} \v S_{\v r} .
\nn\label{eq:full-H-with-anisotropy}\ea
We take a cubic lattice structure rather than the full B20 lattice
structure of MnSi\cite{hopkinson}.

Without the anisotropy terms and the magnetic field, $H_S$ favors the
spiral spin (SS) ground state in which the spiral spins lie in a
plane orthogonal to the propagation vector $\v k$: $S_{\v r} \sim \v
S_{\v k} e^{i \v k\cdot \v r} + \v S_{\v k}^* e^{-i\v k\cdot \v r}$,
$\v S_{\v k} \!\cdot \!\v k = 0$. The $\v k$ stretches along the
$[111]$ ([11] in 2D) direction with the magnitude $k = |\v k|$ fixed
by $\tan k = (K/J)(1/\sqrt{d})$, $d$ being the spatial dimension.
The pitch of the helix can be very long compared to lattice spacing
due to the small $K/J$ ratio in a material. Under high pressure there
is also experimental evidence for the realization of a phase with
multiple ordering vectors\cite{partial-order}. Throughout this paper
we denote such a multiple-$\v k$ spin structure as Skyrme crystal
(SC). The reason for the nomenclature will become transparent
shortly.

Monte Carlo (MC) simulated annealing procedure was employed to work
out the ground states for varying anisotropy strengths $(A_1, A_2)$
and the field strength $H$ oriented along the $z$-direction: $\v H =
H\hat{z}$. Some simplifications were made to save the computational
cost. First, a 2D rather than the 3D lattice was used. Because the
realistic modulation period is very large and difficult to simulate,
we also choose the ratio $K/J = 2\pi/6$ (Hereafter we will take
$J\equiv 1$) which would give $k = 2\pi/6$ in 2D without the
anisotropy. Calculations were mostly carried out for $18\times 18$
lattice, with occasional checks on a $30\times30$ lattice to ensure
consistency. $2\times 10^5$ MC steps were used at each temperature
in the annealing process. It turns out that the same ground state is
found over a widely different choices of $A_1$, and here we present
all the results for $A_1 = 0.5$ without loss of generality. Once the
ground state has been obtained for a given $A_2$ and $H$, we analyze
its structure by making the Fourier transform $\langle \v S_{\v k}
\rangle = \sum_{\v r} \langle \v S_{\v r} \rangle e^{- i \v k \cdot
\v r}$ of the averaged MC configurations $\langle \v S_{\v r}
\rangle$, and looking at the intensity profile $|\langle \v S_{\v k}
\rangle |^2$. A sharper spectral feature is obtained in this way
than by taking the average of the individual intensities, $\langle
|\v S_{\v k}|^2 \rangle$.

\begin{figure}[ht]
\begin{center}
\includegraphics[scale=0.75]{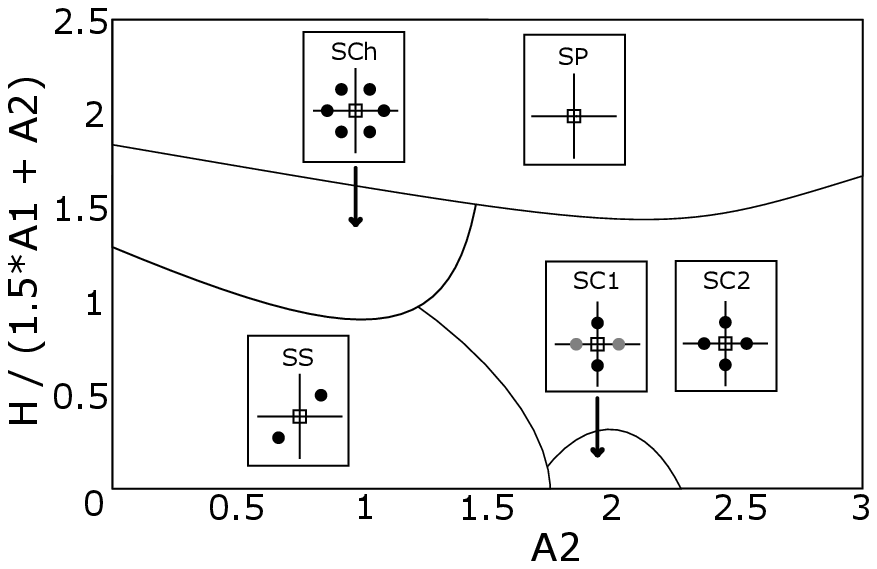}
\end{center}
\caption{(color online)  Low temperature ($T=0.01$) phase diagram of
the spin model in Eq. (\ref{eq:full-H-with-anisotropy}) with
$K=\sqrt{6}$, $A_1 = 0.5$, and $\v H = H\hat{z}$. Phase boundaries
are drawn on the basis of MC simulations at a large number of $(A_2,
H)$ locations. Spin configurations are abbreviated as SS (spiral
spin), and SC (spin crystal). Full spin polarization (SP) results at
high field. The spin crystal phases are further classified as SC$_1$
(unequal Bragg intensities), SC$_2$ (equal Bragg intensities), and
SC$_h$ (hexagonal Bragg spots). The corresponding Bragg patterns are
schematically shown. The Bragg peak at $\v k = 0$ emerges due to the
field-induced uniform magnetization. On the vertical axis $H$ is
divided by an arbitrary energy scale $1.5 A_1 \!+\! A_2$ to ensure a
similar transition temperature to SP for all $A_2$.}
\label{fig:phase-diagram}
\end{figure}

A pair of sharp peaks are obtained at $\pm \v k$ for a single SS,
where $\v k = (k,k)$. Each spin spiral has a right-handed helicity
consistent with the sign of $K>0$. When $A_2$ is sufficiently large,
one finds the SC phase given as the superposition of two pairs of
spin spirals, with modulation vectors at $(\pm k,0)$ and $(0,\pm k)$.
Depending on whether the Bragg intensities $|\langle \v S_{\v
k}\rangle |^2$ are the same or different for the two pairs, we denote
them as SC$_1$ (non-identical) or SC$_2$ (identical). In practice
SC$_1$ is fragile, occupying only a tiny fraction of the phase
diagram. A third class of SC state is found when the field strength
in exceed of a certain threshold value $H_c$ is applied to the SS
phase. This state, denoted SC$_h$, is characterized by three sets of
modulation vectors which are related by 120$^\circ$ rotations. 
We later point out that this phase, and SC$_1$ and SC$_2$ phases
under nonzero magnetic field, also carries a nonzero uniform spin
chirality and anomalous Hall conductivity. The low temperature
schematic phase diagram as obtained from examination of the Bragg
intensity $| \langle  \v S_{\v k}\rangle|^2 $ is shown in Fig.
\ref{fig:phase-diagram}.

Although the details of the phase transitions among the phases
deserves further, more careful study in the future work, we can make
some qualitative statements based on symmetry.  The transition from
SS to SC$_h$ and SS to SC$_1$ are necessarily first-order, while
SC$_1$ $\leftrightarrow$ SC$_2$ can be second-order. In going from
SC$_h$ or SC$_2$ to the fully spin-polarized (SP) state at high
field, the intensity of Bragg peaks gradually diminishes, leaving
only the $\v k =0$ peak in the SP phase.

\begin{figure}[ht]
\begin{center}
\includegraphics[scale=0.45]{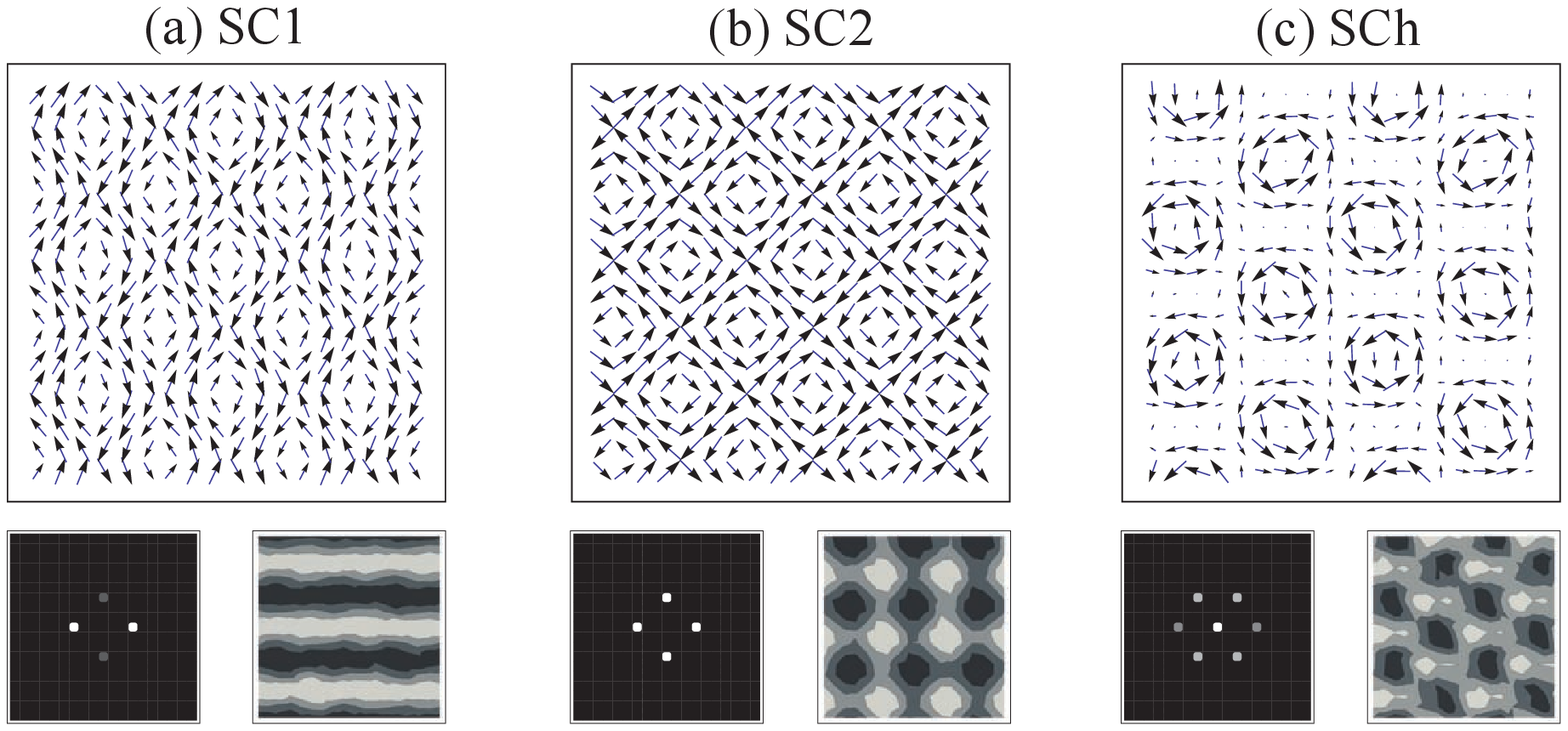}
\end{center}
\caption{(color online)  A plot of the spin configuration projected
on the $xy$ plane $(S^x_i, S^y_i)$ in the three spin crystal ground
states: (a) SC$_1$ at $(A_1, A_2, H) = (0.5, 2.0, 0.0)$, (b) SC$_2$
at $(A_1, A_2, H) = (0.5, 3.0, 0.0)$, and (c) SC$_h$ at $(A_1, A_2,
H) = (0.5, 0.0, 2.0)$. At the bottom left of each figure are the
plots of the Bragg intensity $| \langle \v S_{\v k} \rangle |^2$
showing two (SC$_1$, SC$_2$) and three (SC$_h$) sets of modulation
vectors. Shown at the bottom right are the plots of the local
chirality $\chi_{\v r}$. Bright (dark) regions correspond to
Skyrmions (anti-Skyrmions).} \label{fig:SC}
\end{figure}

The projection of spin patterns onto the $xy$ plane is displayed in
Fig. \ref{fig:SC} for the three Skyrme crystal states. We introduce
the local chirality $\chi_{\v r}$ at the lattice site $\v r$ as

\be  8\pi \chi_{\v r} = \v S_{\v r} \cdot (\v S_{\v
r\!+\!\hat{x}}\times \v S_{\v r\!+\!\hat{y}}) \!+\! \v S_{\v r}
\cdot (\v S_{\v r\!-\!\hat{x}}\times \v S_{\v r\!-\!\hat{y}}) .
\label{eq:lattice-skyrmion-number}\ee
It is well known that a single localized Skyrmion would give the
uniform chirality $\chi = \sum_{\v r} \chi_{\v r}$ equal to unity.
Plots of $\chi_{\v r}$ in Fig. \ref{fig:SC} clearly display the
presence of Skyrmion (bright) and anti-Skyrmion (dark) regions. The
skyrmion density map is largely one-dimensional for SC$_1$,
checkerboard-like for SC$_2$, and hexagonal for SC$_h$. The uniform
chirality $\chi$ was zero for the states obtained at zero field.

\begin{figure}
\begin{center}
\includegraphics[scale=0.75]{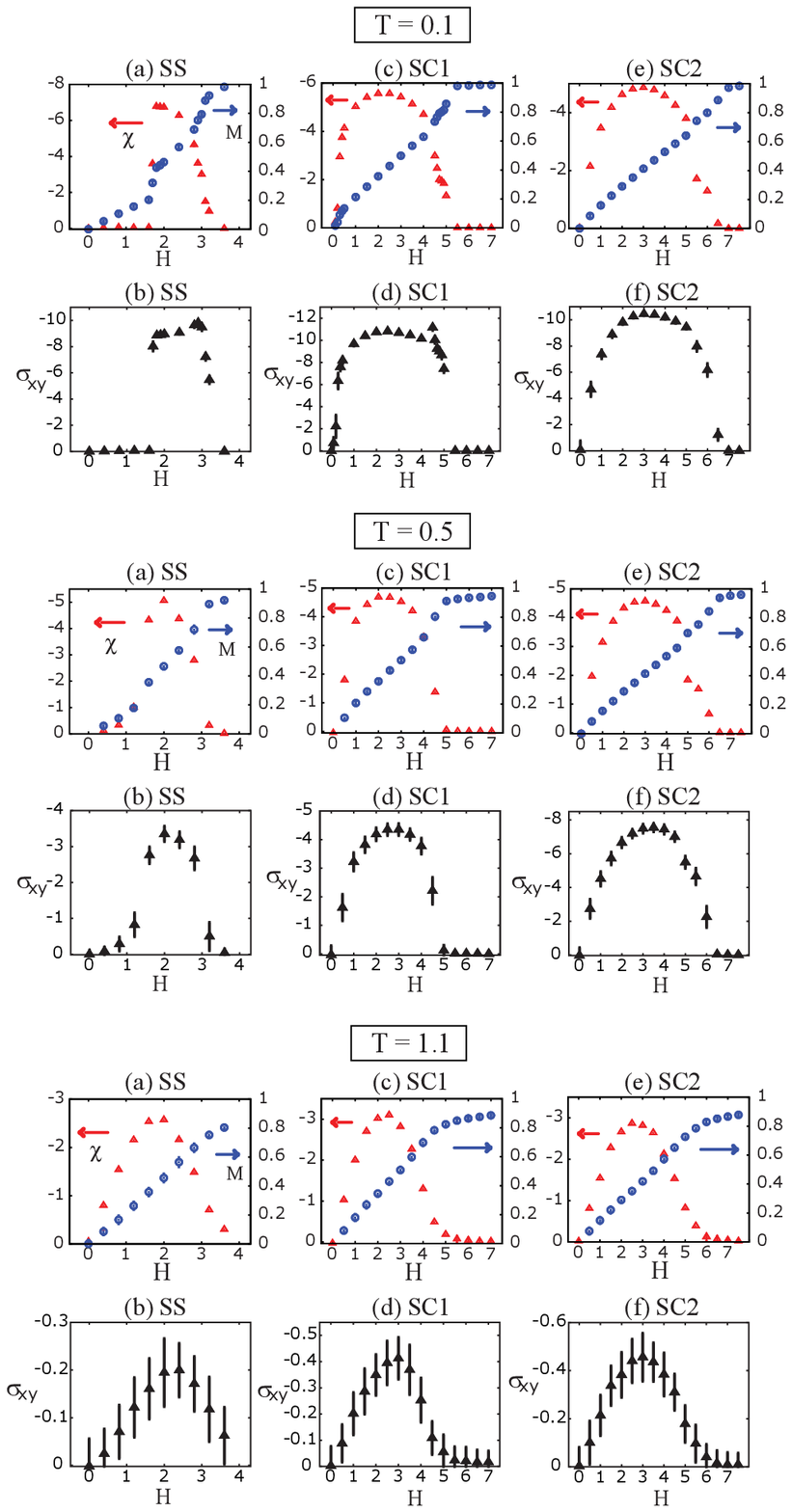}
\end{center}
\caption{(color online) Upper panels: Plot of uniform magnetization
$M$ (blue circle) and uniform chirality $\chi$ (red triangle) with
varying field strength $H$, $\v H = H \hat{z}$, for (a) SS at $(A_1
, A_2 ) = (0.5, 0.0)$, (c) SC$_1$ at $(A_1 , A_2 ) = (0.5, 2.0)$,
and (e) SC$_2$ at $(A_1 , A_2 ) = (0.5, 3.0)$, at three temperatures
$T=0.1$ (top six figures), $0.5$ (middle), and $1.1$ (bottom). Lower
panels: $\langle \sigma_{xy} \rangle$ averaged over 100 MC-generated
spin configurations are shown in (b), (d), and (f) at the
corresponding temperatures and field strengths. Thermal fluctuations
are indicated as error bars.} \label{fig:sigma-on-H}
\end{figure}

Expanding $\v S_{\v r} = \sum_{\v k} \v S_{\v k} e^{i \v k \cdot \v
r }$, the uniform chirality reads  $\chi = \sum_{\v k_1 , \v k_2 }
\v S^*_{\v k_1 + \v k_2} \! \cdot \!\v S_{\v k_1} \!\times \!\v
S_{\v k_2} \cos (k_{1x} \!+\! k_{2y})$.  Our calculation explicitly
shows that this quantity is zero without the magnetic field. With
the field on there appears the $\v k = 0$ component of spin, $\v
S_{\v 0} \propto \v H$. One might think that a non-zero chirality
results from $\chi \sim \v H \cdot \sum_{\v k} \v S_{\v k} \times \v
S_{-\v k}$, but since $\v S_{\v k} \times \v S_{-\v k} \sim i \v k$
and thus $\v S_{\v k} \times \v S_{-\v k} + \v S_{-\v k} \times \v
S_{\v k} = 0$, the chirality contribution from a given spin spiral
is zero. Rather, we find that the uniform chirality is induced in
the SC$_{1}$ and SC$_2$ phases due to the appearance of additional
Bragg peaks at $\v S_{\pm \v k_x \!\pm \! \v k_y}$, $\v k_{x} =
(k,0)$ and $\v k_{y} =(0,k)$, and the nonzero value of the triple
product such as $\v S^*_{\v k_x \! +\! \v k_y} \! \cdot \!\v S_{\v
k_x} \!\times \!\v S_{\v k_y}$. Why the $\v S_{\v k_x + \v k_y}$
component appears with $H$ can be understood from a Ginzburg-Lanau
theory which should contain, on symmetry grounds, a quartic coupling
such as $ (\v S_{\v k_1} \!\cdot\! \v S_{\v k_2} ) (\v S_{\v k_3}
\!\cdot \!\v S_{\v k_4})$. With the appearance of $\v S_0 \sim \v
H$, the coupling becomes $(\v S_{\v k_x} \!\cdot\! \v S_{\v k_y} )
(\v S^*_{\v k_x \!+\! \v k_y}\!\cdot\! \v H)$. Since $\v S_{\v
k_x}\!\cdot\! \v S_{\v k_y}$ is already nonzero in the SC phase, it
implies the appearance of a linear term in $S_{\v k_x \!+\! \v k_y}$
which will lead to the condensation of $S_{\v k_x \!+\! \v k_y}$
order and $\chi$ proportional to $H$. The numerical results shown in
Fig. \ref{fig:sigma-on-H} confirm the linear growth of $\chi$ with
$H$. For SC$_h$, the uniform chirality is found to arise from the
nonzero triplet product $\chi \sim \v S_{\v k_1} \cdot \v S_{\v k_2}
\times \v S_{\v k_3}$, where the three independent modulation
vectors form $\v k_1 + \v k_2 + \v k_3 = 0$.

For field oriented away from the $z$-axis, the Ginzburg-Landau
argument would continue to predict the presence of a linear-$H$
dependence of the chirality in the SC$_1$ and SC$_2$ phases and lack
therefore for the SS phase. Both these predictions are confirmed by
the numerical calculations. Some check was made on the
finite-temperature behavior, using the thermal-averaged intensity
$|\langle \v S_{\v k} \rangle |^2$ was used as a measure of magnetic
ordering. The largest intensity in $|\langle \v S_{\v k} \rangle
|^2$ among all values of $\v k$ decreased continuously until it
reaches zero at the critical temperature. There was no clear
evidence for the existence of an intermediate phase.

Having studied the magnetic phase diagram, we turn to the coupling
of the local moments to the conduction electrons that would result
in the anomalous Hall effect. We adopt the $sd$ Hamiltonian

\be H = -t \sum_{\v r \v r' \sigma} \cbar_{\v r \sigma} c_{\v r'
\sigma} -\lambda \sum_{\v r} \v S_{\v r} \cdot \cbar_{\v r \alpha}
(\sigma )_{\alpha\beta} c_{\v r \beta} , \label{eq:H-sd}\ee
with the moment distribution $\{\v S_{\v r}\}$ obtained from
previous MC calculation. The intrinsic anomalous Hall conductivity
$\sigma_{xy}$ is calculated from

\be \sigma_{xy} = {2\pi \over L^2}\sum_{m \neq n} {f_n \!-\! f_m
\over \eta^2 + (\varepsilon_m \!-\! \varepsilon_n )^2 }\mathrm{Im}
\Big( \langle m |J_x | n \rangle \langle n |J_y | m \rangle \Big)
\label{eq:sigma-xy}\ee
expressed in units of $e^2 /h$. The sum $\sum_{m\neq n}$ extends over
all non-identical pairs of single-particle eigenstates of Eq.
(\ref{eq:H-sd}), $f_m$ is the Fermi function $1/(e^{\beta
\varepsilon_m } \!+\! 1)$, and $L^2$ gives the number of lattice
sites. $J_x$ and $J_y$ are the current operators. The same
temperature $T$ is used both in the Monte Carlo generation of sample
spin configurations, and in evaluating $\sigma_{xy}$. Thermal average
$\langle \sigma_{xy}\rangle$ was taken over 100 MC-generated spin
configurations. The relaxation rate $\eta=0.1$, and the $sd$ coupling
$\lambda =1$ were used in the calculation, with similar results at
other parameter choices. Since the relativistic spin-orbit coupling
term is absent in Eq. (\ref{eq:H-sd}), the nonzero Hall conductivity
we will report below can only be due to the topological Berry phase
effects.

We find $\langle \sigma_{xy} \rangle = 0$ for all the spin
configurations at all temperatures when $H=0$. For finite fields,
the onset of nonzero $\langle \sigma_{xy}\rangle$ and nonzero $\chi$
coincided almost perfectly, confirming the earlier theoretical
anticipation that an unconventional anomalous Hall conductivity
arises as a consequence of uniform chirality $\chi$ being
nonzero\cite{ye,geller,review}. We also verified that the relation
between $\langle \sigma_{xy} \rangle$ and $\chi$ are linear when
both quantities are sufficiently small. Such a close tie between
$\chi$ and $\langle \sigma_{xy} \rangle$ suggests that measurement
of $\sigma_{xy}$ in a given material can be used as an effective
probe of the underlying spin structure. For instance, linear rise of
$\sigma_{xy}$ with the magnetic field would be consistent with the
Skyrme crystal spin structure, but not with the spiral spin
structure.

The field dependence of uniform magnetization $M$, uniform chirality
$\chi$, and $\langle \sigma_{xy}\rangle $ are displayed in Fig.
\ref{fig:sigma-on-H} for several temperatures. Under the full
polarization $M=1$, $\langle \sigma_{xy}\rangle $ and $\chi$
naturally must go to zero, resulting in the characteristic dome shape
of the $\langle \sigma_{xy}\rangle$ curve. A roughly linear
relationship $\sigma_{xy}\!\propto\! H \!\propto\! M$ exists in both
SC$_1$ and SC$_2$ for small $M$. The slope $\sigma_{xy}/H$ decreases
gradually as the temperature rises. The low-temperature $\sigma_{xy}$
can reach up to $\approx 10 e^2 /h$, similar to the value reported in
the model calculation by Metalidis and Bruno\cite{bruno}.

Now we discuss the possible relevance of our results to MnSi.
Several mechanisms for de-stabilizing the SS order in a chiral
magnet have been proposed\cite{tewari,bogdanov,spin-crystal,rosch}.
Tewari \textit{et al.} argued for the existence of a chiral liquid
phase on the high pressure side of MnSi, \textit{i.e.} a
non-magnetic phase with nonzero $\langle \psi_{\v r} \rangle $,
$\psi_{\v r}$ being the chiral order parameter $\psi_{\v r} = \v
S_{\v r} \cdot \nabla \times \v S_{\v r} $. Provided that one can
associate their pressure variable with $A_2$ in our model, we find
the skyrme crystal phase characterized by $\chi_{\v r} = \v S_{\v r}
\cdot (\partial_x \v S_{\v r} \times
\partial_y \v S_{\v r} )$ in place of their chiral liquid order. Robler
\textit{et al}.\cite{bogdanov} considered a Ginzburg-Landau model
which supports a Skyrme crystal phase at the intermediate temperature
range between the low-temperature SS and the high-temperature
paramagnet. Our model in contrast exhibits the skyrme crystal phase
at \textit{low temperature}, driven by the \textit{large anisotropy
term} comparable to $J$ and/or $K$.

Several features of the experiments on MnSi might be understood
within the framework of our model. The first-order change with $A_2$
in going from SS to SC phase and the re-orientation of the primary
ordering vector from [11] to [10] direction in our model are
consistent with the first-order collapse of $T_c$ at the critical
pressure and the re-orientation of the main Bragg peak from [111] to
[110] in MnSi, respectively\cite{partial-order}. The appearance of an
anomalous Hall signal on the high pressure side\cite{ong2} and over a
finite temperature and field segments of the phase diagram might
indicate the realization of a topologically nontrivial spin
structure, according to the close tie between anomalous Hall
conductivity and nonzero spin chirality found in our study. It is
also tempting to associate the recently observed hexagonal Bragg
spots in the so-called A-phase of MnSi\cite{hexagonal-SC} with the
SC$_h$ found in our spin model. Further measurement of the anomalous
Hall conductivity in the same phase will help clarify the connection.
The highly nonlinear threshold behavior of $\sigma_{xy}$ in the SS
phase has not yet been clearly resolved in existing measurements of
ordinary Hall effect by Lorentz force or the anomalous Hall effect on
MnSi\cite{ong1,ong2}.

We close with a caveat that the correspondence between our theory and
experiments on MnSi rests on the physical assumption that increasing
pressure does lead to the increase of $A_2$ over other energy scales.
Whether the Mn orbitals responsible for the localized moments in MnSi
will give rise to such an interaction, and if the pressure does have
an effect on $A_2$ remains a subject for future consideration.
Nevertheless, the predicted field dependence of $\sigma_{xy}$ on the
underlying spin structure is independent of the microscopic spin
model being used.

\acknowledgments  N. N. is supported by Grant-in-Aids under the grant
numbers 16076205, 17105002, 19019004, and 19048015 from the Ministry
of Education, Culture, Sports, Science and Technology of Japan. S.O.
is supported by Grants-in-Aid for Scientific Research (No. 20029006
and No. 20046016) from the MEXT of Japan. H. J. H. is supported by
the Korea Research Foundation Grant (KRF-2008-521-C00085,
KRF-2008-314-C00101) and in part by the Asia Pacific Center for
Theoretical Physics.

\end{document}